\documentclass[10pt,showpacs,preprintnumbers,superscriptaddress]{revtex4}
\usepackage{bbm}
\usepackage[british]{babel}
\usepackage{mathtools}
\usepackage{graphicx,color}

\usepackage[colorlinks,bookmarks=true]{hyperref}
\usepackage[figure,table]{hypcap}
\hypersetup{
	bookmarksnumbered,
	pdfstartview={FitH},
	citecolor={black},
	linkcolor={black},
	urlcolor={black},
	pdfpagemode={UseOutlines}
}

\newcommand{\e}{{\rm e}}
\renewcommand{\i}{{\rm i}}

\begin{document}

\title{Quantum search algorithms on a regular lattice}

\author{Birgit Hein}
\affiliation{School of Mathematical Sciences, University of Nottingham, University Park, Nottingham NG7 2RD, UK} \affiliation{Institute for Theoretical Physics, University of Regensburg, 93040 Regensburg, Germany.}

\author{Gregor Tanner}
\affiliation{School of Mathematical Sciences, University of Nottingham, University Park, Nottingham NG7 2RD, UK} 

\date{20th May 2010}

\pdfbookmark[1]{Abstract}{abstract}

\begin{abstract}
Quantum algorithms for searching one or more marked items on a $d$-dimensional lattice 
provide an extension of Grover's search algorithm including a spatial component. 
We demonstrate that these lattice search algorithms 
can be viewed in terms of the level dynamics near an avoided crossing of a  
one-parameter family of quantum random walks.
We give approximations for both the level-splitting at the avoided crossing and  
the effectively two-dimensional subspace of the full Hilbert space spanning the level crossing. 
This makes it possible to give the leading order behaviour for the search time and the localisation 
probability in the limit of large lattice size including the leading order coefficients. 
For $d=2$ and $d=3$, these coefficients are calculated explicitly.
Closed form expressions are given for higher dimensions.
\end{abstract}

\maketitle

\section{Introduction}
\label{sec:introduction}

Quantum random walks as introduced by Aharonov, Davidovich and Zagury \cite{ADZ93} in 1993 
have gained considerable attention over the last decade or so. It could be demonstrated that 
a quantum version of a classical random walk has transport properties which 
exhibit polynomial or even exponential speed-up compared to a classical 
random walk, see \cite{kempe, Kendon06, Santha08} for overviews. This has recently lead to increased 
efforts for implementing quantum random walks experimentally. In particular, one dimensional 
quantum walks have been realised using neutral atoms in optical lattices \cite{Kar09}, 
with trapped ions \cite{Sch09,Xue09,Zae10}, using coupled optical wave guides 
\cite{Per08} and single photons \cite{Schr10}.  

One of the most fascinating applications based on quantum random walk concepts are spatial 
quantum search algorithms. Like Grover's search algorithm \cite{Gro96, Gro97} for searching an 
unstructured data base, quantum walk search algorithms can (usually) achieve a quadratic 
speed up compared to the corresponding classical search. The most prominent algorithms 
are the search on a hypercube introduced by Shenvi, Kempe and Whaley \cite{SKW03} and 
the search on a $d$-dimensional lattice presented by Childs and Goldstone \cite{CG04} in 
a continuous time version and by Ambainis, Kempe and Rivosh (AKR) \cite{AKR05} as a 
discrete time algorithm.  In particular, it has been pointed out that the search in a 
two-dimensional lattice is critical with the number of search steps scaling like 
${\cal O}(\sqrt N \log N)$ whereas quantum search algorithms on the hypercube as 
well as for lattices of dimension $d\ge 3$ scale like ${\cal O}(\sqrt{N})$; here, 
$N$ is the number of vertices. Expressions for the leading order coefficients for the search 
time in the hypercube have been given in \cite{HT09}, proposals for improving the efficiency 
of the hypercube search can be found in \cite{PGKJ09}. 
In \cite{Tul08}, it has been shown that the number of steps 
for solving the search problem on a two-dimensional lattice can be 
decreased to ${\cal O}(\sqrt{N\log N})$ by modifying the quantum walk search algorithm, 
thus coming closer to the theoretical lower bound $\Omega(\sqrt{N})$
\footnote{The scaling behaviour of a function will be denoted as follows: 
$f\left(x\right)={\cal O}\left(g\left(x\right)\right)$ indicates that there exist two 
positive constants $x_{0}>0$ and $a>0$, such that for all  $x>x_{0}$ the inequality 
$0\le f\left(x\right)\le a g\left(x\right)$ is true. Similarly $f\left(x\right)=
\Omega\left(g\left(x\right)\right)$ will be written if  $0\le b g\left(x\right)\le 
f\left(x\right)$ for all $x>x_{0}$ for some constants $x_{0},\ b>0$. Furthermore, 
$f\left(x\right)=\Theta\left(g\left(x\right)\right)$ denotes that $f\left(x\right)=
{\cal O}\left(g\left(x\right)\right)$ as well as $f\left(x\right)=\Omega\left(g\left(x\right)\right)$.}
\cite{BBBV97}. By changing the starting state, it has been demonstrated in \cite{HT09I} that 
quantum search algorithms on lattices can also be used in a sender-receiver configuration. 
Remarkably, this makes it possible to communicate across the lattice by sending information
exclusively between two (or more) marked vertices where neither the sender nor the receivers need 
to know each others position.   

We will in the following focus on the AKR search algorithm on $d$-dimensional lattices. Extending the ideas and techniques from \cite{HT09}, we will 
give improved estimates for the approximate eigenstate of the walk localised on the target vertex. 
This makes it possible to give closed form expressions for the leading 
order coefficients both for the search time and the search efficiency, that is, the overlap of the localised state with the target vertex. The paper is 
structured as follows: the search algorithm on the lattice and the basic vectors spanning the search space in the full Hilbert space are
introduced in section \ref{chap:grid}; this is followed by a detailed calculation of the normalisation constant for the approximate 
eigenvector localised on the target vertex in section \ref{grid:calculation b}. 
In section \ref{sec:avoidedcrossings}, we analyse the spectral gap at the avoided crossing in an (approximately) invariant two-dimensional
subspace. The results for leading order contributions to the localisation time and the amplitude at the target vertex are given 
in section \ref{grid:results}.

\section{A quantum search on a $d$-dimensional square lattice - introducing the algorithm}
\label{chap:grid}

Quantum search algorithms - like Grover's search - are usually described in terms of a two unitary 
matrices: a quantum or wave propagator acting uniformly on the search space and a ``marker'' 
(or oracle) matrix which deviates from the identity only locally near the marked item. 
We will discuss here a lattice search algorithm of this type which was first introduce in \cite{AKR05}. 
We will limit the discussion to the search of one target vertex only. A generalisation to more than 
one marked item is straightforward and is discussed in \cite{AKR05,HT09I}.
We will first introduce the propagator without a marked vertex 
leading to a quantum random walk on the lattice. The full quantum search 
algorithm and an analysis of the search mechanism will be given in Sec.\ 
\ref{secgrid:search}.\\

\subsection{A quantum random walk on a $d$-dimensional lattice}
\label{secgrid:randomwalk}

We consider a $d$-dimensional regular lattice with $n$ vertices along each dimension, that is, $N=n^{d}$ vertices overall. The positions of the vertices in the lattice are defined by the set of vectors $\left|x\right\rangle=\vec{x}=\left(x_{1}, x_{2},\dots,x_{d}\right)$ with integer coordinates $x_{i}\in\left\{0,1,\dots,n-1\right\}$. Periodic boundary conditions will be assumed throughout. 

Following the spirit of quantum graph theory, we consider quantum waves propagating freely along 1-dimensional bonds connecting adjacent vertices \cite{KS99,GS06}. Waves undergo scattering at the vertices described by a $2d\times 2d$ scattering matrix $\sigma$, where $2d$ is the number of bonds connected at each vertex. The scattering mechanism is the same at all vertices and the quantum dynamics on the lattice is in fact equivalent to a quantum random walk, see \cite{Tan06}. (The unitary scattering matrix $\sigma$ can then also be interpreted as a 'coin flip' matrix acting on an internal (spin-) degree of freedom of a quantum walker). The total Hilbert space $\cal H$ of the problem is described as the tensor product of a vertex or position space $\left|x\right \rangle$ and a direction or coin space $\left |i^\pm \right \rangle$ representing waves travelling in the negative/positive direction along the $i$th axis, $i=1,\ldots, d$. The dimension of $\cal H$ is thus $2d \,n^d = 2d\, N$.

We chose a scattering matrix $\sigma$ which is unbiased with respect to the outgoing directions (except possibly for back-scattering). A natural choice is 
Grover's matrix \cite{Gro97} (which incidentally also features prominently in a quantum graph approach  using Kirchhoff boundary conditions on the vertices, 
see \cite{KS99,GS06}). That is, we define \cite{AKR05}
\begin{align}
\label{defgrid:localcoin}
 \sigma=2\left|s\right\rangle\left\langle s\right| - \mathbbm{1}_{2d}
,\end{align}
where $|s\rangle$ is the uniform distribution in coin space, that is, 
\[
\left|s\right\rangle=\frac{1}{\sqrt{2d}}\sum_{i=1}^{d}\left(\left|i^{+}\right\rangle+\left|i^{-}\right\rangle\right) .
\]
We have identical scattering processes at all vertices and we thus define a global coin flip matrix 

\begin{align}
C=\sigma\otimes\mathbbm{1}_{N} 
.\end{align}

The wave propagation on the lattice is now given by a ``shift'' matrix $S$. Waves emanating from vertex $\left|x\right \rangle$ in the $i^\pm$th direction will reach vertex $\left |x \pm e_i\right \rangle$ with    
$\left|e_i\right\rangle$, the unit vector in the $i^+$th direction.  Writing $\left| i^{\pm} x\right\rangle=\left| i^{\pm}\right\rangle\otimes\left| x\right\rangle$, we find for the shift matrix $S$
\begin{align}
\label{defgrid:S}
S=\sum_{\vec{x}} \sum_{i=1}^{d}\left(\left|i^{+} x-e_{i}\right\rangle\left\langle i^{-} x\right| + \left|i^{-} x+e_{i}\right\rangle\left\langle i^{+} x\right|\right).
\end{align}

This is a natural choice for a moving shift since a walker leaving vertex $\vec{x}$ in direction $i^{+}$, enters vertex $\vec{x}+\vec{e_{i}}$ from direction $i^{-}$.

The quantum random walk $U_{0}$ is defined by first applying the global coin flip and then the moving shift
\begin{align}
\label{introdef:U}
U_0=SC
.\end{align}

In what follows, the eigenvectors and eigenvalues of $U_0$ will be of  importance which have been discussed in some detail in \cite{AKR05}.
Using the tensor product, we write each eigenvector of $U_{0}$ as a vector in coin space $\left| u_{\vec{k}}^{c}\right\rangle$ and a position space vector $\left|X_{\vec{k}}\right\rangle$ according to
\[
\left| \phi_{\vec{k}}^{c}\right\rangle=\left| u_{\vec{k}}^{c} \right\rangle \otimes\left|X_{\vec{k}}\right\rangle
\]
where $\vec{k}$ is a $d$-dimensional vector with components $k_{i}\in\left\{0,1,\dots,n-1\right\}$ and $c = 1,\ldots 2d$.
The vector in position space can be factorised in the form  $\left|X_{\vec{k}}\right\rangle = \bigotimes_{i=1}^{d} \left| \chi_{k_{i}}\right\rangle$, where the $\left| \chi_{k_{i}}\right\rangle$ are obtained from the canonical basis vectors of position space using a Fourier transformation, that is,
\begin{align}
\label{defgrid:chi}
 \left| \chi_{k_{i}}\right\rangle=\frac{1}{\sqrt{n}} \sum_{j=0}^{n-1} \alpha^{k_{i}j}\left|j\right\rangle
 \hspace{1.5cm} {\rm where\ \  }\alpha=\e^{2\pi\i/n}
.\end{align}
The thus obtained basis provides a convenient way to denote the eigenvectors in position space.

For the quantum search, only $2N-1$ of the $2dN$ eigenvectors of $U_0$ are important, namely those having a coin space component not orthogonal to $\left|s\right\rangle$. 
These vectors are the $1$-eigenvector $\left|\phi_{0}\right\rangle=\left|s\right\rangle\otimes\left|X_{\vec{0}}\right\rangle$, 
which is the uniform distribution, and two eigenvectors \cite{AKR05}
\begin{align}
\label{defgrid:eigenvectorU}
\left|\phi_{\vec{k}}^{\pm}\right\rangle=\left|u_{\vec{k}}^{\pm}\right\rangle\otimes\left|X_{\vec{k}}\right\rangle
\end{align}
for each $\vec{k}\neq\vec{0}$ with complex conjugated eigenvalues $\e^{\pm\i\theta_{\vec{k}}}$ and
\[ \cos\theta_{\vec{k}}=\frac{1}{d}\sum_{i=1}^{d}\cos \frac{2\pi k_{i}}{n}. \]

The expressions for the vector $\left|u_{\vec{k}}^{\pm}\right\rangle$ in 
coin space become more and more cumbersome with increasing $d$ and we will not attempt
to give closed form expressions here. However, it has been shown in  
\cite{AKR05} that 
\begin{align}
\label{defgrid:overlap}
\left\langle s\mid\ u_{\vec{k}}^{\pm}\right\rangle=\frac{1}{\sqrt{2}},
\end{align}
which is all we need in what follows. In addition, one has $\left(d-1\right)N+1$ eigenvectors with eigenvalue $1$ and $\left(d-1\right)N$ eigenvectors with eigenvalue $-1$ all with coin components 
perpendicular to $|s\rangle$. Note that the high degeneracies in the $\pm 1$ eigenspaces are due to the special choice of the coin flip matrices $\sigma$ and thus $C$.

\subsection{The quantum search algorithm}
\label{secgrid:search}

\subsubsection{The quantum search matrix}
\label{secgrid:quantsearch}

We now assume that the lattice contains one marked vertex at (a yet unknown) position $\vec{v}$; here, the marking is done by applying a different coin flip or scattering matrix $\sigma^{\prime}$ at this target vertex $\left|v\right\rangle$. The search algorithm is then  defined as
\begin{align}
\label{introdef:Uprime}
U_1=S C'
\end{align}
\begin{align}
\label{defgrid:Cprime}
C^{\prime}=C-\left(\sigma-\sigma^{\prime}\right)\otimes \left|v\right\rangle \left\langle v\right|
.\end{align}
Note that, since $\left| v \right\rangle\left\langle v\right|$ is a projection on the target vertex, 
the additional term changes at most $4d^{2}$ matrix elements in $U_0$. As $U_0$ is a $2dN\times 2dN$-matrix, 
$U_{1}$ is identical to $U_0$  up to a local perturbation and
the quantum search algorithm can be regarded as a locally perturbed quantum random walk.

Following AKR\index{} we define $\sigma^{\prime}=-\mathbbm{1}_{2d}$ which
leads after some algebra to
\begin{align}
\label{defgrid:Uprime}
U_1=U_0\left(1 - 2 \left|sv\right\rangle \left\langle sv\right| \right)
.\end{align}

\begin{figure}[htb]
\centering
\begin{minipage}[t]{0.45\textwidth}
\centering
\includegraphics[scale=0.28]{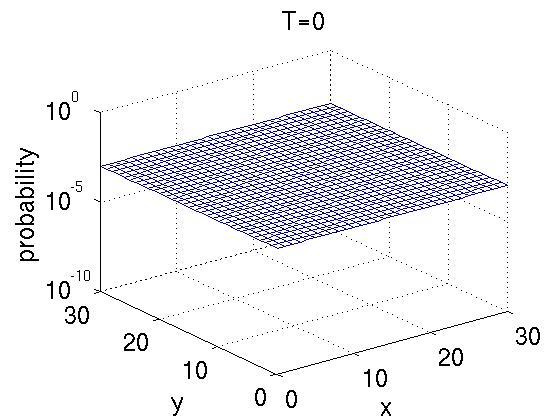}
\end{minipage}
\begin{minipage}[t]{0.45\textwidth}
\centering
\includegraphics[scale=0.28]{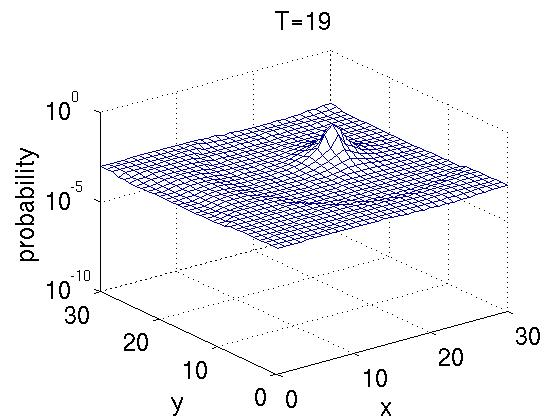}
\end{minipage}
\begin{minipage}[t]{0.45\textwidth}
\centering
\includegraphics[scale=0.28]{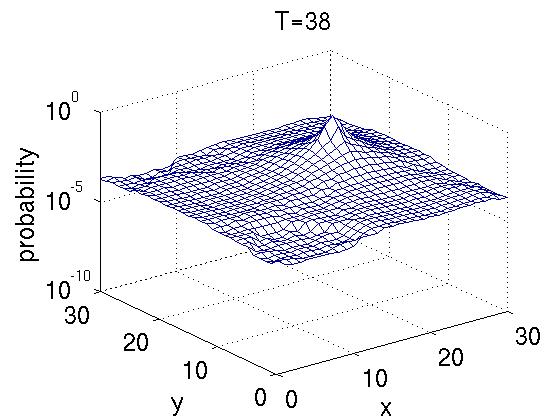}
\end{minipage}
\begin{minipage}[t]{0.45\textwidth}
\centering
\includegraphics[scale=0.28]{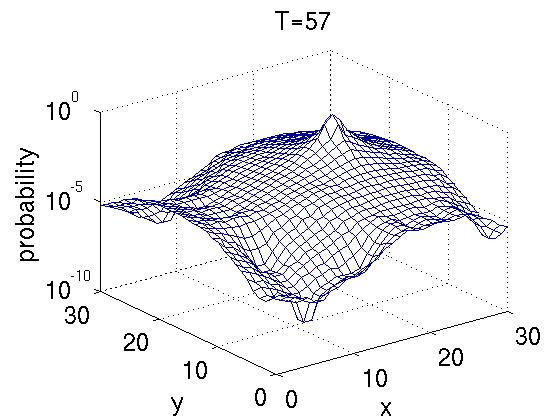}
\end{minipage}
\caption{Probability distribution of the quantum walk on a $31\times 31$-lattice for a number of time steps $t$ (on a logarithmic scale).}
\label{figgrid:search}
\end{figure}

The perturbed quantum walk is shown in Fig.\ \ref{figgrid:search};
starting from the uniformly distributed state $\left|\phi_{0}\right\rangle$ and applying 
$U_1$ for $t$ time steps with $t=0,\ 19,\ 38$ and $57$, one observes a localisation on the 
target vertex. Plotted here is the probability (wave amplitude square) on a logarithmic scale. \\

Following the analysis developed for the hypercube in \cite{HT09}, we now define a one parameter family of unitary matrices $U_{\lambda}$ for the $d$-dimensional lattice:
\begin{align} 
\label{defgrid:U_lambda}
U_{\lambda}= U_0+\left(\e^{\i\pi\lambda}-1\right) U_0\left|sv\right\rangle\left\langle sv\right|
.\end{align}

This makes it possible to extrapolate smoothly from the unperturbed walk ($\lambda = 0$) to the AKR search algorithm ($\lambda = 1$) and will be helpful for analysing the eigenvalues and eigenvectors  of $U_\lambda$ near $\lambda = 1$ in terms of the eigenbasis of the unperturbed quantum walk.

Before doing so, we simplify the problem by reducing the size of the Hilbert space.  We note that 
eigenvectors of $U_0$ with eigenvalues $\pm 1$ (except the important vector $\left|\phi_{0}\right\rangle$) are orthogonal to $\left|sv\right\rangle$ and thus remain eigenvectors of $U_\lambda$  independent of the value of $\lambda$; 
this part of $\cal H$ is thus irrelevant for the localisation effect. We define a reduced space ${\cal H}^{\prime}$ by projecting out all $\pm 1$-eigenvectors of $U$ 
except $\left|\phi_{0}\right\rangle$. This subspace is a $2N-1$ dimensional Hilbert space spanned by  $\left|\phi_{0}\right\rangle$ and  $|\phi_{\vec{k}}^{\pm}\rangle$ for $\vec{k}\neq\vec{0}$.

The scalar product  $\langle\phi_{\vec{k}}^{\pm}\mid sv\rangle$ can be obtained 
using (\ref{defgrid:overlap}); thus an expansion of $\left| sv\right\rangle$ in terms of eigenvectors of $U_{0}$ in the reduced space gives
\begin{align}
\left| sv\right\rangle
\label{defgrid:svinHprime}
&=\frac{1}{\sqrt{N}} \left|\phi_{0}\right\rangle +\frac{1}{\sqrt{2N}}\sum_{\vec{k}\neq\vec{0}} \alpha^{-\vec{k}\vec{v}}\left(\left|\phi_{\vec{k}}^{+}\right\rangle+\left|\phi_{\vec{k}}^{-}\right\rangle\right)
.\end{align}
Since $\left| sv\right\rangle$ is orthogonal to all eigenvectors of $U_0$ not contained in 
$\cal{H}^{\prime}$, this expansion also holds for the non-reduced space.


\subsubsection{Approximate eigenvectors of $U_{\lambda}$}

\begin{figure}
\centering
\includegraphics[scale=0.35]{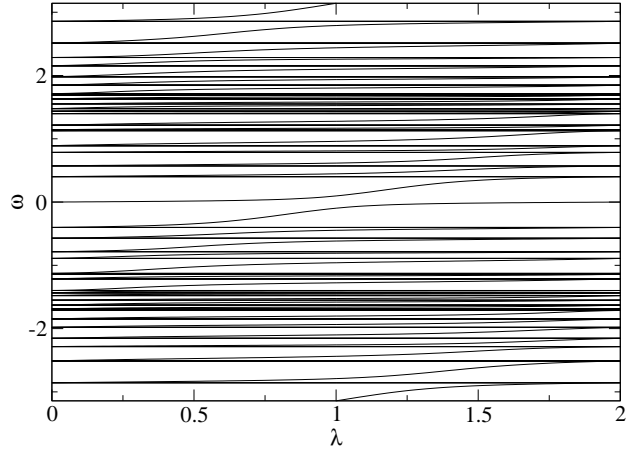}
\caption{The eigenphases as functions of $\lambda$ for $n=11$, $d=2$.}
\label{figgrid:eigenphases}
\end{figure}
In the following we will analyse the search algorithm in detail. Our approach is based on the work in \cite{HT09} and makes it possible to go beyond the results found in \cite{AKR05} by giving explicit leading order coefficients for the search time. 

The overall strategy is best explained by considering briefly the 
spectrum of eigenphases $\{ \omega_i\}$ of $U_{\lambda}$; in 
Fig.\ \ref{figgrid:eigenphases}, the eigenphases are shown for 
an $11\times 11$ lattice in the reduced space ${\cal H}^{\prime}$. The 
local perturbation leads to a set of avoided crossings in the spectrum, 
most notably at $\lambda=1,\omega=0$. 
This avoided crossing is formed by the state $\left|\phi_{0}\right\rangle$ 
(which is a 1-eigenstate of U) and ''another'' state, which we will 
denote $\left| \nu_\lambda\right \rangle$ in what follows. It will be argued 
that $\left| \nu_\lambda\right \rangle$ is localised in the vicinity of 
the target vertex and that the subspace $\{ \left|\phi_{0}\right\rangle, 
\left| \nu_\lambda\right \rangle\}$ is an approximate basis for the 
two-dimensional eigenspace spanned by the exact eigenstates at the crossing. 
The quantum search amounts then to a rotation in this two-dimensional 
subspace of ${\cal H}'$ similar to the mechanism found in 
Grover's original search. It is important to note that unlike for Grover's
algorithm, for the lattice search only approximate expressions for the 
state $|\nu_{\lambda=1}\rangle$ are known which will be given below; 
furthermore, the starting state $\left|\phi_{0}\right\rangle$ has a small 
overlap with the rest of the spectrum at $\lambda = 1$ leading to losses.

We start by considering the uniform distribution $\left|\phi_{0}\right\rangle$ which is the remaining $1$-eigenvector of $U_0$ in ${\cal H}'$. One obtains
\begin{align}
\label{grid:Ulambdaphi0}
U_{\lambda}\left|\phi_{0}\right\rangle
&=\left|\phi_{0}\right\rangle+\left(\e^{\i\lambda\pi}-1\right) U_0\left|sv\right\rangle \frac{1}{\sqrt{N}} 
,\end{align}
that is, $|\phi_0\rangle$ is an approximate eigenvector of $U_\lambda$ of order ${\cal O} \left(N^{-1/2}\right)$. The remaining components, $U_{0}\left|sv\right\rangle$, are localised 
on the $2d$ sites next to the target vertex. This can be seen from the definition of the shift operator $S$ in Eq.\ (\ref{defgrid:S}).

To obtain $\left| \nu_\lambda\right \rangle$, we start by expanding the 
vector in the basis of the unperturbed walk $U_0$, that is,
 \begin{align}
\label{def:nu}
\left|\nu_{\lambda}\right\rangle= a_{0}\left|\phi_{0}\right\rangle +\sum_{\vec{k}\neq\vec{0}}\left(a_{\vec{k}}^{+}\left|\phi_{\vec{k}}^{+}\right\rangle+a_{\vec{k}}^{-}\left|\phi_{\vec{k}}^{-}\right\rangle\right)
\end{align}
with a yet unknown set of coefficients $\{a_{0},a_{\vec{k}}^{\pm}\}$. Normalisation requires
\begin{align}
\left|a_{0}\right|^{2}+\sum_{\vec{k}\neq\vec{0}}\left(\left|a_{\vec{k}}^{+}\right|^{2}+\left|a_{\vec{k}}^{-}\right|^{2}\right)=1.
\end{align}

We now make the ansatz 
\begin{align} \label{app-eigen}
U_{\lambda}\left|\nu_{\lambda}\right\rangle\approx\e^{\i g\left(\lambda\right)}\left|\nu_{\lambda}\right\rangle,
\end{align}
that is, we assume an approximate eigenvalue equation with a yet unknown eigenvalue $\e^{\i g\left(\lambda\right)}$. This gives rise to a linear system of equations for the $2N-1$ unknown coefficients $\{a_0, a^\pm_{\vec{k}}\}$. We are looking for a second approximate eigenvector which spans the space at the avoided crossing together with $\left|\phi_0\right\rangle$, that is 
we demand that $\left|\nu_{\lambda}\right\rangle$ and $\left|\phi_{0}\right\rangle$ are orthogonal which leads to $a_{0}=0$.

A straight forward application of $U_{\lambda}$ on $\left|\nu_{\lambda}\right\rangle$ results in
\begin{align} \label{soso}
U_{\lambda}\left|\nu_{\lambda}\right\rangle
=&\sum_{\vec{k}\neq\vec{0}}\left(a_{\vec{k}}^{+}\e^{\i\theta_{\vec{k}}}\left|\phi_{\vec{k}}^{+}\right\rangle+a_{\vec{k}}^{-}\e^{-\i\theta_{\vec{k}}}\left|\phi_{\vec{k}}^{-}\right\rangle\right)  +\left(\e^{\i\pi\lambda}-1\right)\nonumber \\
 & \cdot \left(\frac{1}{\sqrt{N}} \left|\phi_{0}\right\rangle +\frac{1}{\sqrt{2N}}\sum_{\vec{k}\neq\vec{0}} \alpha^{-\vec{k}\vec{v}}\left(\e^{\i\theta_{\vec{k}}}\left|\phi_{\vec{k}}^{+}\right\rangle+\e^{-\i\theta_{\vec{k}}}\left|\phi_{\vec{k}}^{-}\right\rangle\right)\right)\left\langle sv\mid \nu_{\lambda}\right\rangle 
\end{align}
with $\vec{\nu}$, the position of the target vertex. In what follows, we set
\begin{align} 
\label{defgrid:b}
b:=\left\langle sv\mid \nu_{\lambda}\right\rangle
\end{align}
and define the overall phase factor of $\left|\nu_{\lambda}\right\rangle$ such that $b$ is real 
with $0\le b\le 1$. Note that $b$ determines the overlap of the approximate eigenvector
with the target vertex. We will show in the next section that $b$ is of order 
${\cal O}(1/\sqrt{\ln N})$
for $d=2$ and ${\cal O}(1)$ for $d\ge 3$ at $\lambda = 1$; thus $|\nu_1\rangle$ is indeed localised at the target vertex.

We now add and subtract the right hand side of Eq.\ (\ref{app-eigen}) in (\ref{soso}) and obtain
\begin{align}
U_{\lambda}\left|\nu_{\lambda}\right\rangle
=& \e^{\i g\left(\lambda\right)}\left|\nu_{\lambda}\right\rangle +\frac{b\left(\e^{\i\pi\lambda}-1\right)}{\sqrt{N}} \left|\phi_{0}\right\rangle 		\nonumber \\
  +&  \sum_{\vec{k}\neq\vec{0}}\Bigg(\left(a_{\vec{k}}^{+}\left(\e^{\i\theta_{\vec{k}}}-\e^{\i g\left(\lambda\right)} \right)+\frac{b\left(\e^{\i\pi\lambda}-1\right)\alpha^{-\vec{k}\vec{v}}\e^{\i\theta_{\vec{k}}}}{\sqrt{2N}}
  \right)\left|\phi_{\vec{k}}^{+}\right\rangle \nonumber \\
  & +\left(a_{\vec{k}}^{-}\left(e^{-\i\theta_{\vec{k}}}-\e^{\i g\left(\lambda\right)} \right)+\frac{b\left(\e^{\i\pi\lambda}-1\right)\alpha^{-\vec{k}\vec{v}}\e^{-\i\theta_{\vec{k}}}}{\sqrt{2N}}
  \right)\left|\phi_{\vec{k}}^{-}\right\rangle\Bigg) 
.\end{align}
An approximate  solution is obtained if each of the coefficients in front of the components $\left|\phi_{\vec{k}}^\pm\right\rangle$ is zero, that is,
\begin{align}
\label{grid:a_k}
a_{\vec{k}}^{\pm}=\frac{b\left(\e^{\i\pi\lambda}-1\right)\alpha^{-\vec{k}\vec{v}}\e^{\pm\i\theta_{\vec{k}}}}{\sqrt{2N}\left(\e^{\i g\left(\lambda\right)} -e^{\pm\i\theta_{\vec{k}}}\right)}.
\end{align} 
The resulting vector $U_\lambda \left|\nu_\lambda\right \rangle$ has a component in the 
$|\phi_0 \rangle$ direction of order ${\cal O}(1/\sqrt{N})$; thus, similar to Eq.\  (\ref{grid:Ulambdaphi0}), $|\nu_\lambda \rangle$ remains in a subspace spanned by itself and $|\phi_0\rangle$.
 
The coefficients $a_{\vec{k}}^{\pm}$ enter Eq.\ (\ref{grid:a_k}) also through $b$, that is, (\ref{grid:a_k}) represents a set of linear, coupled equations. Assuming $b\neq 0$, equation (\ref{defgrid:b}) can be divided by $b$ and the system of equations has a solution if and only if
\begin{align}
\label{grid:sumformula}
1 & =\frac{\left(\e^{\i\pi\lambda}-1\right)}{2N}\sum_{\vec{k}\neq\vec{0}}\left( \frac{\e^{\i\theta_{\vec{k}}}}{\e^{\i g\left(\lambda\right)} -e^{\i\theta_{\vec{k}}}}
+\frac{\e^{-\i\theta_{\vec{k}}}}{\e^{\i g\left(\lambda\right)} -e^{-\i\theta_{\vec{k}}}}
\right)
.\end{align} 

The avoided crossing occurs at $\lambda=1$ and $\e^{\i g\left(1\right)}=1$, for which the right hand side of equation (\ref{grid:sumformula}) can be obtained directly, 
\begin{align}
\label{grid:sumformula_result}
\frac{-2}{2N}\sum_{\vec{k}\neq\vec{0}}\left( -1\right) 
=\frac{1}{N}\left(N-1\right)=1-\frac{1}{N}
.\end{align} 
That is, no solution exists which maps a vector $|\nu_1\rangle$ exactly onto itself and  $|\phi_{0}\rangle$. 
The error term ${1}/{N}$ is small compared to typical coupling terms between basis states $|\phi_{\vec{k}}^\pm \rangle$ (which are of the order $1/\sqrt{N}$)
and the set of coefficients from Eq.\ (\ref{grid:a_k}) thus define an approximate solution of the eigenvalue equation (\ref{app-eigen}). That is, the vector 
\begin{align}
\label{grid:nu_lambda}
\left|\nu_{\lambda}\right\rangle
=&\frac{b\left(\e^{\i\pi\lambda}-1\right)}{\sqrt{2N}} \nonumber\\
&\sum_{\vec{k}\neq\vec{0}}\alpha^{-\vec{k}\vec{v}}	
 \cdot \left(\frac{\e^{\i\theta_{\vec{k}}}}{\e^{\i g\left(\lambda\right)} -e^{\i\theta_{\vec{k}}}}\left|\phi_{\vec{k}}^{+}\right\rangle
+\frac{\e^{-\i\theta_{\vec{k}}}}{\e^{\i g\left(\lambda\right)} -e^{-\i\theta_{\vec{k}}}}\left|\phi_{\vec{k}}^{-}\right\rangle\right)
,\end{align}
\footnote{This vector $\left|\nu_{\lambda=1} \right\rangle$ plays essentially the same role as the vector $\left|\omega_{good}\right\rangle$ considered by AKR \cite{AKR05}. Both vectors are in good approximation in the subspace spanned by the two exact eigenvectors at the crossing. However, the vector in \cite{AKR05} is not orthogonal to $\left|\phi_{0}\right\rangle$ and one needs information about the exact eigenphases of $U_{\lambda=1}$ at the crossing. For the vector presented here, a diagonalisation of  $U_{\lambda=1}$ is not necessary.} 
fulfils to leading order the equation 
\begin{align}
\label{grid:Ulambdanu}
U_{\lambda}\left|\nu_{\lambda}\right\rangle
&= \e^{\i g\left(\lambda\right)}\left|\nu_{\lambda}\right\rangle +\frac{b\left(\e^{\i\pi\lambda}-1\right)}{\sqrt{N}} \left|\phi_{0}\right\rangle 
\end{align}
where $g(\lambda)$ is defined implicitly by minimising the expression (\ref{grid:sumformula}). An expansion of $g(\lambda)$ around $\lambda = 1$ can be obtained successively, see \cite{HT09}. We will restrict our attention here to the case $\lambda = 1$ and thus $\left|\nu_{\lambda=1} \right\rangle$ 
for which  $g(1) = 0$.

The only unknown quantity in determining $\left|\nu_{1}\right\rangle$ is the constant $b$ which has dropped out of the equations. Since the vector $\left|\nu_{1}\right\rangle$ needs to be normalised, $b$ turns out to be a normalisation constant.

\section{Normalisation of  $|\nu_1\rangle$}
\label{grid:calculation b}

Evaluating the normalisation constant $b$ turns out to be the most laborious and technical part of the derivation. Readers more interested in the final results may want to proceed directly the result in equation (\ref{grid:1/b2}).

We will in the following restrict ourselves to the case 
$\lambda=1$ and thus $\e^{\i g\left(1\right)}=1$ only. Demanding the vector in Eq. (\ref{grid:nu_lambda}) to be normalised, one obtains 
\begin{align}
\frac{1}{\left|b\right|^{2}}
&=\frac{4}{2N} \sum_{\vec{k}\neq\vec{0}} \frac{2}{\left|1 -e^{\i\theta_{\vec{k}}}\right|^{2}}  \\
&=\frac{2d}{N} \sum_{\vec{k}\neq\vec{0}} \frac{1}{d -\sum_{i=1}^{d}\cos \frac{2\pi k_{i}}{n}}
,\end{align}
where the sum is taken over all vectors $\vec{k}\neq\vec{0}$ in a $d$-dimensional cube with  $k_{i}\in\{0,1,\dots ,n-1\} $. 

We start by rearranging the total sum into a summation over lower dimensional objects where only a limited number of entries of $\vec{k}$ are different from $0$. 
That is, we consider the summation over all one dimensional edges, two dimensional faces, $3$-dimensional cubes and so on, where the non-zero entries $k_{i}$ vary from $1$ to $n-1$. The edges are obtained by choosing $d-1$ entries of $\vec{k}$ equal $0$, the faces have $d-2$ entries of $\vec{k}$ equal $0$ and higher dimensions accordingly. This new arrangement results in
\begin{align}
\frac{1}{\left| b\right|^{2}} 
=& \frac{2d}{N} \sum_{\vec{k}\neq\vec{0}} \frac{1}{d -\sum_{i=1}^{d}\cos \frac{2\pi k_{i}}{n}} \\
=&\frac{2d}{N}  \underbrace{d}_{\text{number of edges}}  \underbrace{\sum_{j_{1}=1}^{n-1}\frac{1}{1 -\sum_{l=1}^{1}\cos \frac{2\pi j_{l}}{n}}}_{\text{contribution of one edge}} \nonumber \\
 & +\frac{2d}{N}  \underbrace{{d \choose 2}}_{\text{number of faces}}  \underbrace{\sum_{j_{1},j_{2}=1}^{n-1}\frac{1}{2 -\sum_{l=1}^{2}\cos \frac{2\pi j_{l}}{n}}}_{\text{contribution of one face}} \nonumber \\
 &+ \dots  \\
\label{defgrid:1/b^2-sumformula}
=
&\frac{2d}{N}  \sum_{i=1}^{d}{d \choose i} \sum_{j_{1},j_{2},\dots,j_{i}=1}^{n-1}\frac{1}{i -\sum_{l=1}^{i}\cos \frac{2\pi j_{l}}{n}}  \\
\label{defgrid:defI_i}
=& \frac{2d}{N}\sum_{i=1}^{d}{d \choose i}\left(\frac{n}{\pi}\right)^{i} I_{i} 
,\end{align}
where  $I_{i}$ has been implicitly defined in (\ref{defgrid:defI_i}).
Using the identity $1-\cos x=2\sin^{2}\frac{x}{2}$, we can simplify the sums and obtain 
\begin{align}
I_{i}=\frac{1}{2}\left(\frac{\pi}{n}\right)^{i} \sum_{j_{1},\dots,j_{i}=1}^{n-1}\left(\sum_{l=1}^{i}\sin^{2}\frac{\pi j_{l}}{n}\right)^{-1}
.\end{align}

Using Poisson's summation formula, one arrives at 
\begin{align}
I_{i}=&\frac{1}{2}\left(\frac{\pi}{n}\right)^{i} \int_{\frac{1}{2}}^{n-\frac{1}{2}}dx_{1} \sum_{m_{1}=-\infty}^{\infty}\e^{2\pi\i m_{1}x_{1}}\dots	\nonumber \\
 & \int_{\frac{1}{2}}^{n-\frac{1}{2}}dx_{i} \sum_{m_{i}=-\infty}^{\infty}\e^{2\pi\i m_{i}x_{i}}    \left(\sum_{l=1}^{i}\sin^{2}\frac{\pi x_{l}}{n}\right)^{-1}  \\ \nonumber
=&\frac{1}{2}\int_{\frac{\pi}{2n}}^{\pi-\frac{\pi}{2n}}dy_{1} \dots \int_{\frac{\pi}{2n}}^{\pi-\frac{\pi}{2n}}dy_{i}	\nonumber 		\\ \label{sums}
 &\sum_{m_{1}=-\infty}^{\infty}\dots \sum_{m_{i}=-\infty}^{\infty}  \frac{\e^{2\i n\left(m_{1} y_{1}+\dots +m_{i} y_{i}\right)}}{\sum_{l=1}^{i}\sin^{2} y_{l}}
,\end{align}
where the last equality is due to a reordering of terms and substituting $y_{j}=\frac{\pi}{n}x_{j}$ in all integrals.

The dominant contributions come from the terms $m_{j} = 0$  for all $j=1,\ldots i$, that is, we write to leading order in the large $n$ limit,
\begin{align} \label{integrals1}
I_{i}&=\frac{1}{2}\int_{\frac{\pi}{2n}}^{\pi-\frac{\pi}{2n}}dy_{1} \dots \int_{\frac{\pi}{2n}}^{\pi-\frac{\pi}{2n}}dy_{i}  \frac{1}{\sum_{l=1}^{i}\sin^{2} y_{l}}
.\end{align}
Using the symmetry of the sine squared function, the integrals (\ref{integrals1}) can be simplified to
\begin{align}
\label{grid:I_i}
I_{i}&=2^{i-1}\int_{\frac{\pi}{2n}}^{\frac{\pi}{2}}dy_{1} \dots \int_{\frac{\pi}{2n}}^{\frac{\pi}{2}}dy_{i}  \frac{1}{\sum_{l=1}^{i}\sin^{2} y_{l}}.
\end{align}
We are only interested in leading order contributions and will thus use the integral  (\ref{grid:I_i})
as an approximation for the full set of sums in Eq.\ (\ref{sums}). The 
expression (\ref{grid:I_i}) will form the starting point of our calculation of the leading order behaviour of $|b|$.

\subsection{Integration of $I_{1}$}
\label{secgrid:first_integration}

The first integration can be done explicitly and one obtains
\begin{align}
I_{1}=\int_{\frac{\pi}{2n}}^{\frac{\pi}{2}}dy \frac{1}{\sin^{2} y}
= \cot \frac{\pi}{2n} =\frac{2n}{\pi}+ {\cal O}\left(\frac{1}{n}\right)
\label{defgrid:I_1}
.\end{align}

\subsection{Integration of  $I_{2}$}
\label{grid:b-for-d2}

To obtain a leading order estimate for
\begin{align}
I_{2}&=2\int_{\frac{\pi}{2n}}^{\frac{\pi}{2}}dx  \int_{\frac{\pi}{2n}}^{\frac{\pi}{2}}dy \frac{1}{\sin^{2} x+\sin^{2} y} 
\end{align}
we first note that the integrand is symmetric with respect to exchanging $x$ and $y$ and therefore
\begin{align}
I_{2}&=4\int_{\frac{\pi}{2n}}^{\frac{\pi}{2}}dx  \int_{x}^{\frac{\pi}{2}} dy \frac{1}{\sin^{2} x+\sin^{2} y} 
.\end{align}

Now, the $y$ integration is solved by observing that 
\begin{align}
\frac{d}{dx} \frac{\arctan\left(\frac{\tan y}{\tan x}\sqrt{2\tan^{2}x+1}\right)}{\sin x \sqrt{\sin^{2}x+1}}=\frac{1}{\sin^{2} x+\sin^{2} y}
.\end{align}
Thus
\begin{align}
I_{2}&=4\int_{\frac{\pi}{2n}}^{\frac{\pi}{2}}dx  \left[\frac{\arctan\left(\frac{\tan y}{\tan x}\sqrt{2\tan^{2}x+1}\right)}{\sin x \sqrt{\sin^{2}x+1}}\right]_{y=x}^{\frac{\pi}{2}} \\
&=4\int_{\frac{\pi}{2n}}^{\frac{\pi}{2}}dx  \frac{\frac{\pi}{2}-\arctan\left(\sqrt{2\tan^{2}x+1}\right)}{\sin x \sqrt{\sin^{2}x+1}}
.\end{align}
Integrating by parts results in
\begin{align}
I_{2}=&4 \ln \left(\sqrt{2+\cot^{2} \frac{\pi}{2n}}+\cot \frac{\pi}{2n}\right)\left(\frac{\pi}{2}-\arctan\sqrt{2\tan^{2}\frac{\pi}{2n}+1}\right) \nonumber \\
  & -4\int_{\frac{\pi}{2n}}^{\frac{\pi}{2}}dx\frac{\ln \left(\sqrt{2+\cot^{2} x}+\cot x\right)}{\sqrt{2+\cot^{2}x}}
.\end{align}
Using the substitution $z=\tan x$, the resulting integration can finally be written in the form\footnote{The following result was obtained using \textsc{mathematica}.}
\begin{align}
\label{defgrid:I_2}
I_{2}&=\pi\ln n +\pi\ln\frac{4}{\pi}-2K-\frac{\pi}{2}\ln 2+{\cal O}\left(\frac{1}{n^{2}}\right)
,\end{align}
where $K\approx 0.916$ is Catalan's constant.

Using Eq.\ (\ref{defgrid:defI_i}), the result for the $d=2$ dimensional lattice is 
\begin{align}
\frac{1}{\left| b\right|^{2}} =& \frac{4}{N}\sum_{i=1}^{2}{2 \choose i}\left(\frac{n}{\pi}\right)^{i} I_{i}  \\
\label{defgrid:normalisation2d}
=&\frac{2}{\pi} \ln N + \frac{8}{\pi^{2} }(2 - K)  +\frac{2}{\pi} \ln\frac{8}{\pi^2}+{\cal O}\left(\frac{1}{N}\right)
.\end{align}
%

\subsection{Integration of $I_{3}$}
\label{grid:b-for-d3}

For the third integral $I_3$, we will only evaluate the asymptotic limit for $n\to\infty$. Starting with
\begin{align} \label{I3}
I_{3}&=4\int_{\frac{\pi}{2n}}^{\frac{\pi}{2}}dx  \int_{\frac{\pi}{2n}}^{\frac{\pi}{2}}dy  \int_{\frac{\pi}{2n}}^{\frac{\pi}{2}}dz \frac{1}{\sin^{2} x+\sin^{2} y+\sin^{2} z}, 
\end{align}
we evidently have $I_3 >0$ independent of $n$. An upper bound for $I_3$ can be obtained using spherical 
coordinates, that is
\begin{align} 
I_3 < 
4\int_{\frac{\pi}{2n}}^{\frac{\pi}{2}\sqrt{3}}dr  \int_{0}^{\frac{\pi}{2}}d\varphi 
 \int_{\frac{\pi}{2}}^{\pi}d\theta \frac{r^2 \sin\theta}{\sin^{2} x+\sin^{2} y+\sin^{2} z}. 
\end{align}
The integrand is bounded in the whole region of integration and in particular also in the limit $r \to 0$; 
we thus have $I_3 = {\cal O}(1)$. 

Using numerical integration methods, we obtained
\begin{align}
\label{defgrid:I_3}
\lim_{n \to \infty}I_{3}=15.672\dots
.\end{align}


Substituting again into Eq.\ (\ref{defgrid:defI_i}), we obtain for the normalisation of the $d=3$ 
dimensional lattice:
\begin{align}
\frac{1}{\left| b\right|^{2}} 
&=\frac{6}{N}\sum_{i=1}^{3}{3 \choose i}\left(\frac{n}{\pi}\right)^{i} I_{i} 	\\
\label{defgrid:ordernormalisation3d}
&=\frac{6}{\pi^{3}}I_{3}	
=\Theta\left(1\right)
.\end{align} 

\subsection{Integrations for $d>3$}
\label{grid:b-for-d>3}

We show in this section that $1/b^2 = {\cal O}(1)$ for all $d\ge 3$; actual numerical values can be obtained by integrating expressions of the form (\ref{grid:I_i}) using numerical methods.  
 From this, we can deduce that the search time $T$ indeed scales like 
$T \sim \sqrt{N}$ for $d\ge 3$ as will be shown in Sec.\ \ref{sec:avoidedcrossings}.

We will proceed by induction: from  (\ref{defgrid:ordernormalisation3d}) it is evident that ${1}/{\left|b\right|^{2}}={\cal O}\left(1\right)$ holds for $d=3$. Thus, assuming that ${1}/{\left|b\right|^{2}}=
{\cal O}\left(1\right)$ holds for some $d\ge 3$, it follows from Eq.\
(\ref{defgrid:defI_i})
and $I_{i}> 0$, that $I_{d}={\cal O}\left(1\right)$.

It remains to be shown that ${1}/{\left|b\right|^{2}}={\cal O}\left(1\right)$ also holds for $d+1$. 
The starting point is again provided by Eq. (\ref{defgrid:defI_i}), but this time for $d+1$, that is,
\begin{align}
 \frac{1}{\left|b\right|^{2}}= 2\left(d+1\right)\left(\sum_{i=1}^{d}{{d+1} \choose i}\frac{n^{i-d-1}}{\pi^{i}} I_{i}+\frac{1}{\pi^{d+1}} I_{d+1}\right)
.\end{align}
The sum of the first $d$ terms adds to a leading order of ${1}/{n}$, since $I_{i}={\cal O}\left(1\right)$ for  $3\le i\le d$ and the lower order terms $I_{2}=\Theta\left(\ln N\right)$ and $I_{1}=\Theta\left(n\right)$ have prefactors $n^{1-d}$ and $n^{-d}$, respectively. Therefore these contributions vanish for $n\to \infty$.

It remains to be shown that $I_{d+1}={\cal O}\left(1\right)$ and this is done using Eq.\  (\ref{grid:I_i}), that is,
\begin{align}
 I_{d+1}&=2^{d}\int_{\frac{\pi}{2n}}^{\frac{\pi}{2}}dy_{1} \dots \int_{\frac{\pi}{2n}}^{\frac{\pi}{2}}dy_{d+1}  \frac{1}{\sum_{l=1}^{d}\sin^{2} y_{l}+\sin^{2} y_{d+1}}
.\end{align}
As the squared sines are greater than zero, an upper bound is obtained by dropping the last sine term. Now, the $y_{d+1}$ integration is performed, that is, 
\begin{align}
 I_{d+1}&\le
2^{d}\left(\frac{\pi}{2}-\frac{\pi}{2n}\right)\int_{\frac{\pi}{2n}}^{\frac{\pi}{2}}dy_{1} \dots \int_{\frac{\pi}{2n}}^{\frac{\pi}{2}}dy_{d}  \frac{1}{\sum_{l=1}^{d}\sin^{2} y_{l}} 	\\
&=\left(\pi-\frac{\pi}{n}\right) I_{d}
\end{align}
and this results in $I_{d+1}={\cal O}\left(1\right)$. Thus, the normalisation constant $b$ is of the order 
${\cal O}\left(1\right)$.

The overall result for the normalisation constant is to leading order
\begin{align}
\label{grid:1/b2}
\frac{1}{b^2}=
\begin{cases}
\frac{2}{\pi} \ln N + \frac{8}{\pi^{2} }(2 - K)  +\frac{2}{\pi} \ln\frac{8}{\pi^2} & \text{for $d=2$}\\
\frac{6}{\pi^3}I_{3} &\text{for $d=3$} \\
{\cal O}\left(1\right)	&\text{for $d\ge 4$}
\end{cases},
\end{align}
where $K\approx 0.916$ is Catalan's constant and we found $I_{3}\approx15.672$ asymptotically, see (\ref{defgrid:I_3}). 
In particular, it has been shown that for $i\ge 3$, $\lim_{n\to\infty} I_{i}$ converges. Thus, the leading order contribution to $b$ can be calculated from equation (\ref{grid:I_i}) by replacing $\frac{\pi}{2n}$ with $0$ and using numerical integration methods.

\subsection{Localisation}
The normalisation parameter $b$ is at the same time a measure for the localisation of $|\nu_1\rangle$
 onto the state $|sv\rangle$ and thus on the target vertex $\vec{v}$, see Eq.\  (\ref{defgrid:b}). 
Hence, the probability that the search will be localised at the target vertex $\left|v\right\rangle$ is proportional to 
\begin{align}
\left|\left\langle\nu_{1}\mid sv\right\rangle\right|^{2}=b^2
.\end{align}
(In fact, $b^2$ gives only a lower bound for the localisation probability as $\left|s\right\rangle$ fixes the coin space state in which the localised state is measured.
). 

For estimating the overlap of the two-dimensional eigenspace at the 
avoided crossing with the basis vector pair
$|\phi_0\rangle, |\nu_1\rangle$, we can resort to the results,  Eqs.\ 
(\ref{grid:Ulambdaphi0}),
(\ref{grid:Ulambdanu}). In particular, we found in (\ref{grid:Ulambdaphi0}), that the uniform state 
$|\phi_0\rangle$ is mapped onto itself and a component in the direction $U_{0} |sv\rangle$.
One finds by straightforward calculation
\begin{align}
\left\langle \nu_{1}\mid U_0\mid sv\right\rangle
&=- \frac{2b}{2N} \sum_{\vec{k}\neq\vec{0}}
\left(\frac{1}{\e^{-\i g\left(1\right)}-e^{\i\theta_{\vec{k}}} }
+\frac{1}{ \e^{-\i g\left(1\right)}-e^{-\i\theta_{\vec{k}}} }\right)	\\
  &= -b\left(1-\frac{1}{N}\right)
.\label{grid:vUsv}
\end{align}
The overlap of $|\nu_1\rangle$ with $U_{0}\left|sv\right\rangle$, that is, the localisation on the nearest neighbours is thus of the same order as 
on the target state $|sv\rangle$ itself. 
To obtain the last equality $\e^{\i g\left(1\right)}=1$ has been used. 

We have shown that the vector $|\nu_1 \rangle$ constructed in the preceding section 
is localised on the target vertex. We have furthermore shown that the vector space spanned by the orthonormal pair of approximate eigenvectors
$\{|\phi_0, |\nu_1 \rangle\}$ has an order ${\cal O}(1)$ or ${\cal O}(1\sqrt{\ln N})$ 
overlap with itself under the unitary map $U_1$. 

\section{Quantum search at the avoided crossing}
\label{sec:avoidedcrossings}

The quantum search algorithm corresponds to a rotation in a lower 
dimensional sub-space 
spanned by the eigenvectors at the avoided crossing at 
$\lambda = 1, \omega = 0$. The search is initialised in the state 
$\left|\phi_{0}\right\rangle$
which has an ${\cal O}(1)$ overlap with the exact eigenstates at the crossing, see (\ref{grid:Ulambdaphi0}). We have constructed a second
approximate eigenvector at the crossing, namely $\left|\nu_{\lambda = 1}\right\rangle$ in (\ref{grid:nu_lambda}), again with an ${\cal O}(1)$
overlap with the eigenspace at the avoided crossing.
We can thus restrict the analysis to studying the avoided crossing in the 
subspace of $\cal H$ spanned by the basis vectors 
$\{\left|\phi_{0}\right\rangle, \left|\nu_{1}\right\rangle\}$; this 
subspace is approximately invariant under the quantum walk $U_1$, as shown
in Sec.\ \ref{secgrid:search},  see also \cite{HT09,HT09I}.
 
\subsection{The size of the gap at the avoided crossing}
In a first step, we project out the two-dimensional submatrix related to 
the (approximately) invariant subspace, 
namely $\left(U_{1}^{2 \times 2}\right)=\e^{-\i H}$, 
where $H$ is  a Hermitian $2 \times 2$ matrix.
The entries of $H$ can be determined by calculating the matrix elements of $\left(U_{1}^{2 \times 2}\right)$ explicitly. 
Using Eqs.\  (\ref{defgrid:svinHprime}), (\ref{grid:Ulambdaphi0}) and (\ref{grid:Ulambdanu}),
one obtains to leading order
 \begin{align}
 \left\langle\phi_{0}\mid U_{1}\mid\phi_{0}\right\rangle
  &= 1-\frac{2}{N}                              \\
   \left\langle\nu_{1}\mid U_{1}\mid\nu_{1}\right\rangle
  &= 1
\end{align}
 for the diagonal elements and
\begin{align}
 \left\langle\phi_{0}\mid U_{1}\mid \nu_{1}\right\rangle
 &= \frac{-2b}{\sqrt{N}}                        \\
  \left\langle\nu_{1}\mid U_{1}\mid \phi_{0}\right\rangle
 &= \frac{-2}{\sqrt{N}}\left\langle\nu_{1}\mid U_0\mid sv\right\rangle            \\
  &= \frac{2b}{\sqrt{N}}\left(1-\frac{1}{N}\right)
\end{align}
for the off-diagonal entries, where Eq.\ (\ref{grid:vUsv}) has been used 
for the last equality. Identifying the basis vectors $\{\left|\phi_{0}\right\rangle, \left|\nu_{1}\right\rangle\}$ with $\{\left|1\right\rangle, \left|2\right\rangle\}$, we obtain to leading order
\begin{align}
\label{defgrid:U2x2}
U_{1}^{2 \times 2}
=\begin{pmatrix} 1 & -\epsilon  \\ \epsilon &1  \end{pmatrix}
\quad \mbox{with} \quad  
  \epsilon=\frac{2b}{\sqrt{N}}
 \end{align}
which results in a Hamiltonian
 \begin{align} 
\label{grid:H2x2}
H=\begin{pmatrix} 0 & -\i\epsilon \\ \i\epsilon& 0 \end{pmatrix}
.\end{align}

We note that the coupling term $\epsilon$ scales like $1/\sqrt{N}$ and 
depends on the normalisation constant $b$.

Eigenvectors and eigenvalues of $H$ are easily calculated, that is,
\begin{align}
\label{cube:eigenvaluesH}
\begin{matrix}
{\rm eigenvector}&\hspace{0.5cm} &\rm{eigenvalue}\\
 \left|u_{1}\right\rangle=\frac{1}{\sqrt{2}}\begin{pmatrix} 1 \\ i \end{pmatrix} & &-\epsilon \\
  \left|u_{2}\right\rangle=\frac{1}{\sqrt{2}}\begin{pmatrix} 1 \\ -i \end{pmatrix} & &\epsilon 
.\end{matrix}
\end{align}
This means in particular that the gap at the avoided crossing behaves like 
\[\Delta = 2 \epsilon = \frac{4b}{\sqrt{N}}+{\cal O}\left(\frac{1}{N}\right).\]

The time of search is directly related to the gap $\Delta$; 
the search starts in the 
vector $|\phi_0\rangle = 1/\sqrt{2}(|u_1\rangle + |u_2\rangle)$ which 
after successive iterations becomes
\begin{align}
\left(U_{1}^{2 \times 2}\right)^{t}\left| \phi_0\right\rangle
&=\e^{-\i H t}\left|\phi_{0}\right\rangle	=\e^{-\i H t}	\frac{1}{\sqrt{2}}\left( \left|u_{1}\right\rangle+ \left|u_{2}\right\rangle\right)	\\
&=\frac{1}{\sqrt{2}}\left( \e^{\i \epsilon t}\left|u_{1}\right\rangle+\e^{-\i \epsilon t} \left|u_{2}\right\rangle\right)		
.\end{align}
Thus, for a time $T:=\frac{\pi}{2\epsilon} = \frac{\pi}{\Delta}$, 
one has $\e^{\pm\i\epsilon t}=\pm\i$ and
\begin{align}
\left(U_{1}^{2 \times 2}\right)^{t}\left| \phi_0\right\rangle
=\frac{\i}{\sqrt{2}}\left( \left|u_{1}\right\rangle- \left|u_{2}\right\rangle\right)	
=-\left|\nu_{1}\right\rangle
,\end{align}
which is the localised state. Note, that the search time $T$ is inversely proportional to the 
coupling parameter $\epsilon$ in Eq.\ (\ref{grid:H2x2}) and thus to the 
spectral gap at the avoided crossing. The quantum search 
algorithm succeeds after  ${\cal O}(\sqrt{N}/b)$ steps and thus faster as any 
classical search.\\

\section{Results} 
\label{grid:results}
Putting everything together, we have now identified the search time in 
leading order as
 \begin{align}
 \label{defgrid:time}
 T=\frac{\pi}{2\epsilon}=\frac{\pi\sqrt{N}}{4b} .
 \end{align}

It depends on the normalisation constant $b$ which has been obtained in 
(\ref{grid:1/b2}). Overall, the number of time steps $T$ can now be given 
as the integer closest to
 \begin{align}
 \label{grid:time}
T=
\begin{cases}
\frac{\pi\sqrt{N}}{4}\sqrt{\frac{2}{\pi} \ln N +\frac{8}{\pi^2}(2-K)+\frac{2}{\pi}
\ln\frac{8}{\pi^2}} &\text{for $d=2$} \\
\frac{\pi\sqrt{N}}{4} \sqrt{\frac{6 I_{3}}{\pi^{3}}}&\text{for $d=3$} \\
\Theta\left(\sqrt{N}\right)	&\text{for $d\ge 4$.}
\end{cases}	
\end{align}
\begin{figure}
\centering
\includegraphics[scale=0.25]{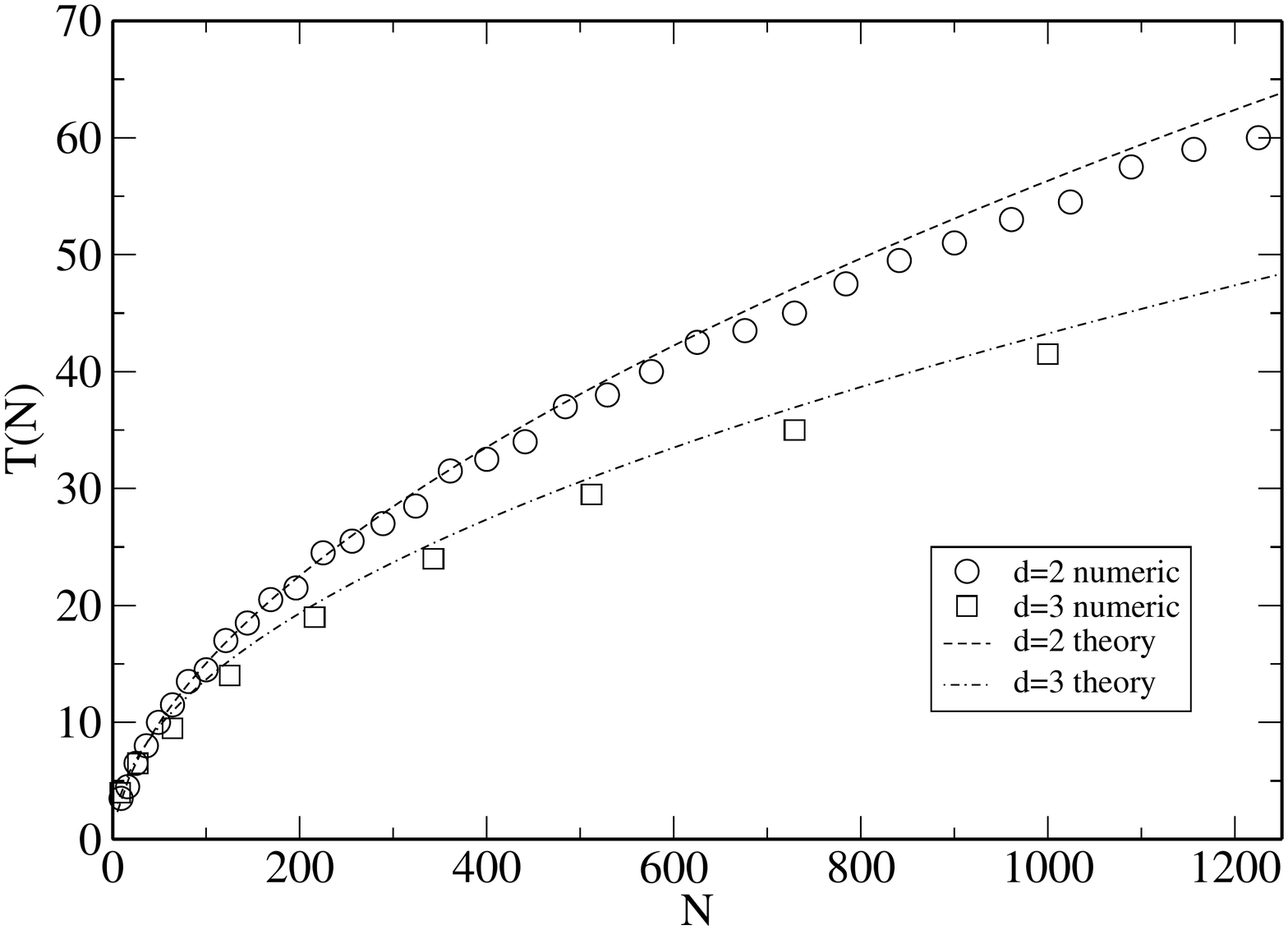}
\caption{The localisation time  $T$ after which the walk is localised at the marked vertex for several $N$. Numerical result compared to analytical for $d=2$ and $3$.}
\label{figgrid:time}
\end{figure}

Note, that the bound for $d\ge 3$ is tight, because we know from 
\cite{BBBV97,Gro96}, 
that the search can not be faster than $\sqrt{N}$.
In Fig.\ \ref{figgrid:time}, we compare the analytical results for the 
localisation time with numerical simulations. The theoretical results 
agree very well with the outcome of the numerical simulations. The general behaviour suggests 
that the walk for a fixed number of vertices is the faster the higher the 
dimension. 

The probability to find the search at the target vertex after T steps is 
equal to $b^{2}$ (\ref{defgrid:b}).  From Eq.\ (\ref{grid:1/b2}), we find

\begin{align}
\label{grid:b}
b=
\begin{cases}
\frac{1}{\sqrt{\frac{2}{\pi} \ln N + \frac{16}{\pi^{2} } +\frac{4}{\pi} \ln\frac{4}{\pi}-\frac{8K}{\pi^{2}}-\frac{2}{\pi}\ln 2+{\cal O}\left(\frac{1}{N}\right)}}=\Theta \left(\frac{1}{\sqrt{\ln N}}\right) &\text{for $d=2$} \\
\frac{1}{\sqrt{\frac{6}{\pi^3}I_{3}}}=\Theta \left(1\right) &\text{for $d=3$} \\
\Theta\left(1\right)	&\text{for $d\ge 4$}
\end{cases}
\end{align}
Note, that the upper bound for $b$ in the case $d\ge 4$ is a consequence of (\ref{defgrid:time}) and the lower bound for the search time is provided by \cite{BBBV97}.

To leading order, the localisation amplitude at the target vertex is not 
$N$-dependent for $d\ge 3$; the $d=2$ case is exceptional, as the projection of 
$\left|\nu_{1}\right\rangle$ onto $\left|v\right\rangle$ decreases like 
${1}/{\sqrt{\ln N}}$. Note, however, that the state 
$|\nu_{1}\rangle$ is still localised on the marked vertex for $d=2$ since 
the average amplitude of an eigenstate of the random walk $U_{0}$ on an arbitrary vertex 
decreases like $\frac{1}{\sqrt{N}}$. Amplitude amplification methods can be used to increase the success probability to a constant in $N$ by repeating the search algorithm ${\cal O}\left(\sqrt{\ln N}\right)$ times \cite{Grover98}.

\begin{figure}[htb]
\centering
\includegraphics[scale=0.30]{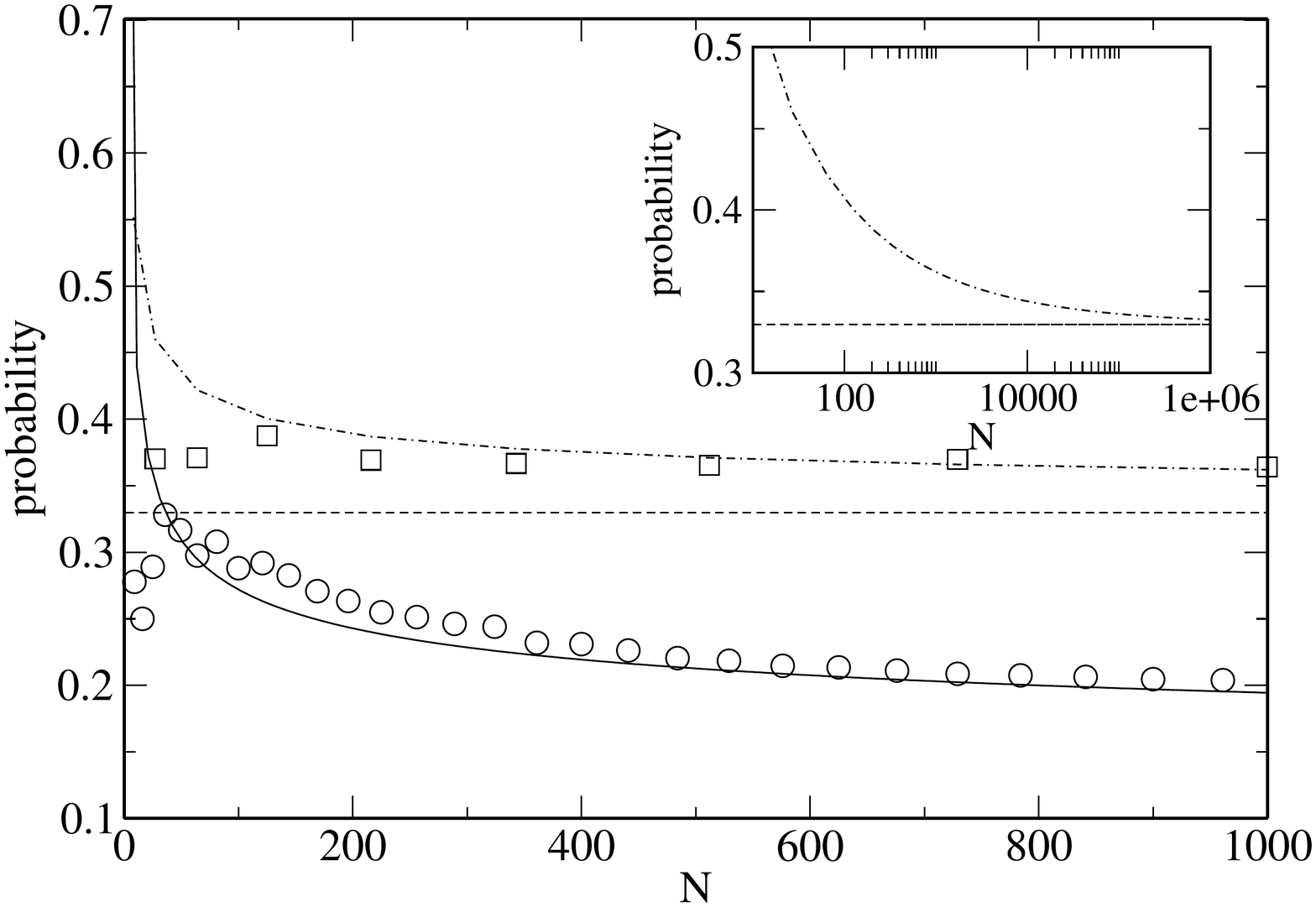}
\caption{Maximal probability at the target vertex as 
a function of $N$: circles ($d=2$) and squares ($d=3$) are
obtained directly from the quantum search; the solid and dashed lines 
represent Eq.\ (\ref{grid:b}).  The dashed-dotted line
gives the result for $b^2$ obtained directly from Eq.\ (\ref{defgrid:defI_i}) 
for $d=3$.  Inset: Direct evaluation of Eq.\ (\ref{defgrid:defI_i}) 
compared with asymptotic result Eq.\ (\ref{grid:b}) for $d=3$ over large 
$N$ interval.}
\label{figgrid:probability}
\end{figure}

In  Fig.\ \ref{figgrid:probability}, we compare the numerical results for the 
probability to find the walk at the target vertex 
with the analytical results for 
$b^2$. 
The numerical results are obtained from 
running the quantum search algorithm and determining 
the maximum of the probability at the target vertex.
It is noted that for $d=3$, the asymptotic value given in 
Eq.\ (\ref{grid:b}) is not reached  before $N \approx 10^6$, 
(see inset of Fig.\ \ref{figgrid:probability}). This corresponds to a 
3d lattice of side-length $n = 100$ - far beyond what can 
be simulated on a computer.  However, a direct evaluation of $b^2$ in terms 
of the sums in Eq.\ (\ref{defgrid:defI_i}) confirms the basic 
mechanism behind our approach.

\section{Conclusions}
\label{sec:conclusions}

We presented a detailed analysis of the lattice search and show in particular 
that the search mechanism is based on a rotation from a uniformly extended 
state to a localised state coupled at an avoided crossing point in the 
spectrum of the operator $U_{\lambda}$. The search time and localisation
strength can be obtained from an effectively two-dimensional model similar
to the analysis in \cite{HT09}. We have thus independently verified the 
scaling behaviour of the search time and localisation probability 
as stated in \cite{AKR05} and can now also give explicit expressions for the 
leading order coefficients; the results have been verified by numerical 
simulations.

A generalisation to a search with more than one marked item is straightforward 
and leads to reduced models as presented in \cite{HT09I}. We also 
give a detailed derivation of how to obtain the localised state
$\left|\nu_\lambda\right\rangle$ asymptotically. This is important for the wave communication 
protocols proposed in \cite{HT09I}; here, the search algorithm is used 
to search for more than one marked vertex, to send signals through the 
lattice from a sender to one or more receiver points and to construct 
new search algorithms that do not rely on knowing the number of target 
vertices.

\pdfbookmark[0]{References}{references}

\end{document}